\documentclass[11pt,a4paper]{article}
\usepackage{amsmath}
\usepackage{amsfonts}
\usepackage{amssymb}
\usepackage{xspace}
          \usepackage{graphics}
\usepackage{graphics}

\renewcommand\th{\theta}
\def\l{\left}  \def\r{\right}
\def\a{\alpha} \def\b{\beta}

\newcommand\proof{\noindent {\sc Proof:}\qquad}
\newcommand\qed{\hfill$\quad${\rule{3mm}{3mm}}\medskip\\}

\newcommand\LB[1]{\label{#1}} 
\newcommand\BE[2]{\begin{#1} #2 \end{#1}}

\newcommand\EQ[2]{\BE{equation}{\LB{#1} #2}}

 \newcommand\EQn[1]{\BE{equation*}{ #1}}

\newcommand{\eq}{equation\xspace}

\newcommand\f{\varphi}
\newcommand\om{\omega}

\newcommand\Th{\Theta}

\newcommand\dl{\delta}

\newcommand{\iy}{\infty}

\def\proof{\noindent {\sc Proof:}\qquad}

\title{Singularity Propagation for the  Gurtin-Pipkin  \eq
}
\author{
{S. A. Ivanov\thanks{St. Petersburg Institute of Terrestrial Magnetism,
Ionosphere and Radio Wave Propagation.
{{\tt sergei.a.ivanov@.mail.ru }}}}
\thanks{
The work was supported by Russian Foundation for Basic Research, RFBR
Project 11-01-00790a and RFBR
Project 11-01-00667a. }
}

\begin{document}
\date{}

\maketitle

\begin{abstract}
We show  that the Dirac delta  function in the boundary condition of the
Gurtin-Pipkin \eq generates a
moving delta-function with an exponentially decreasing factor.
\end{abstract}

\vskip1cm

\section{Introduction}

Gurtin and Pipkin in \cite{GuPip} have derived the first time the equation for
the heat transfer with finite propagation speed in contrast yo the "classical"
heat \eq., see \cite{FW} where this fact has been was first proved.

Now systems of such type are considered in several fields of physics such as
systems with thermal memory \cite{EMV}, viscoelasticity problems \cite{CMD},
and acoustic waves in composite media \cite{Sham}.

We consider the equation of the first order in time on $(0,\iy)$ with zero
initial data
\begin{equation}\label{1}
\theta_t=\int_0^t k(t-s) \theta_{xx} (s)d s  \,,\qquad \theta(x,0)=0
\end{equation}
and with nonhomogeneous DBC
$$
\th(0,t)=u(t),
$$
where $u$ may be a distribution. We suppose that $k>0$. In application $k$ is a
decreasing function.

In the case $k$ is the Dirac delta $\dl(t)$ the \eq becomes the heat one
with the infinite velocity. In  the 'opposite' case
$k(t)=Const$  the \eq \eqref1 is, in a fact, an integrated wave \eq.
In \cite{BK}, \cite{B} the kernel of the form $a/(t+\om)^l$, $l>1$ is considered, what gives
a non-autonomous telegraph-equation, which has a self-similar solution.
The damped wave \eq ( or a kind of the telegraph-equation) arises, if we take an exponential
$k(t)=a\exp(-bt)$, see below.

In a sense in this notes we refine the border separating the system with finite and infinite
velocity of propagation. In the terms of the Laplace transform the dynamical system has a
finite velocity if the image of the kernel devreasing as $ 1/z$ or faster.

Note that in \eqref1 the  integral term gives a stronger
perturbation than in the system of the second order in time.
Also study the spectrum of the systems
corroborate this fact, see \cite{EI},  \cite{RSV}. See also papers about
regularity of solutions,
\cite{I2013}, \cite{RSV}.

The main result of the notes is that
for a  positive  smooth kernel  the main singularity of
the solution generated by $\dl(t)$ in the boundary condition has the form
\EQ{th0}{
e^{-\b x}\dl(t-x/a)
}
with
\EQ{ab}{
a=\sqrt{k(0)}, \ \b=-k'(0)/2k(0).
}
The proof is based on the Laplace
transform of the solution and the Paley - Wiener theorem.

The main term \eqref{th0} coincides with the main term of singularity
of the  damped wave \eq
\EQ{wave1}{
\th_{tt}=a^2\th_{xx}-b\th_t, \ b=2\a\b.
}
Also the regularity and partially behavior  of the solutions to this \eq
is the same as regularity of the solutions to \eqref1, see \cite{I2013}.
This allows us to say that the \eqref1 is  a perturbation of the \eqref{wave1}.

Probably, there are some
relation with the following result. In \cite{P2012} it was proved that
the sharp control time for second order in time  \eq with memory is the same as for
corresponding telegraph \eq.

\section{\LB{main}{Main results}}

\subsection {Finite velocity}

We will work with distribution here and
suppose that the kernel and the solution does not grow too fast and we can apply the
Laplace transform. Apply this  transform to \eqref1 and  denote
the Laplace-image by capital characters.

\BE{theorem}{\LB{finite}
Let $u(t)=\dl(t)$ and
\EQ{K0}{
K(z)=\frac{a^2}z+O\l(\frac{1}{z^2}\r).
}
Then the solution propagates with the velocity $a$.
}
\proof

After Laplace transform we obtain the family of  ODEs
\EQ{new1}{
z\Th(x,z)=K(z)\Th_{xx}(x,z), \ x>0, \ \Th(0,z)=U(z).
}
For every $z$ this differential (in $x$) \eq has constant coefficients. Set
$$
\f(z)=\sqrt{z/K(z)}
$$
(the main branch).
The  solution do not increasing exponentially in the right half plane is
$$
\Th(x,z)=U(z)e^{-\f(z)x}.
$$
Let the boundary condition be the Dirac delta: $u(t)=\dl(t)$.

On the case of \eqref{K0}
$$
\f(z)=\frac za +O(1).
$$
We see that for fixed $x$ the function $e^{xz/a} \Th(x,z)$
is bounded in the right half plane.
By the  Paley-Wiener theorem for distributions, see \cite{Hermander}, Th 7.3.1,
this gives that the preimage $\th(x,t)$ is t-supported on $(0,x/a )$. In the other words,
the front of the solution is moving with the velocity $a$.  \qed

\BE{example}{
The idea of the proof is very clear for the wave equation, i.e., for the case
 $k(t)=a^2$. Evidently, the solution is
$\th =\dl(t-x/a)$. In the Laplace-images we have
$$
K(z)=\frac{a^2}z, \ \f(z)=\frac za.
$$
Therefore the solution to \eqref{new1} is
$$
\Th(x,z)=e^{-\frac xa z}.
$$
The function $e^{xz/a} \Th(x,z)$ is 1 here.
}

\BE{example}{
If $k=1/t^\a$, $\a<1$, then
$$
K=z^{1-\a}, \ \f=z^{1-\a/2}
$$
and the velocity is infinite.
}

\subsection {Propagation of singularities}.

\BE{theorem}{\LB{terms}
Let $u(t)=\dl(t)$ and
\EQ{K}{
K(z)=\frac{a^2}z-\frac{\b}{z^2}+\frac{c}{z^3}+o\l(\frac{1}{z^3}\r), a>0.
}
Then the solution has the form
\EQ{th}{
\th(x,t)=
e^{-bx/2a}\dl(t-x/a)+  p(x)H(t-x/a)+q(x,t),
}
and $q(x,t)$ is a
function continuous  near the front $t=x/a$.
}
\proof  If we have
\EQ{1/z}{
F(z)=e^{-\a z}\l(a_1+a_2\frac1z+a_3\frac1{z^2}+\dots,  \r),
}
then the preimage is
$$
f(t)=a_1\dl(t-\a) +a_2H(t-\a)+a_3(t-\a)_++\dots.
$$

The idea of the proof is to obtain the expansion \eqref{1/z} for $e^{-\f(z)x}$.
For simplicity we omit calculation of the coefficient
$p$ by the Heaviside function.  Find, using \eqref K,
the coefficients in the
expression of $\f$
$$
\f(z)=a_1+a_2\frac1z+O\l(\frac1{z^2}\r).
$$
We have
$$
\frac z{K(z)}=\frac{z^2}{a^2}\,\frac1{1-\b/a^2z+O(1/z^2)}.
$$
From here
$$
\f=\frac za\l[ 1+\frac{\b}{2a^2}\frac1z+  +  O\l(\frac1{z^2}\r)  \r].
$$
Now
$$
\Th(x,z)=e^{-\f(z)x}=e^{-\frac za x}e^{-\frac{\b x}{2a^3}}e^{O(1/z)}
=e^{-\frac za x}e^{-\frac{\b x}{2a^3}}(1+O(1/z)).
$$

Then the preimage of the main term of $\Th$ is
$$
e^{-\frac{\b x}{2a^3}}\dl(t-x/a)=e^{-b x}\dl(t-x/a),
$$
with $b=\b a^3/2$.
\qed

\subsection{ damped wave \eq}

Here we find the singularity in the case of $k$ is an exponential.
If $k(t)=a^2e^{-bt}$ (and $K(z)=\a^2/(z+b)$),
then the differentiation gives a damped (or telegraph)wave \eq \eqref{wave1}.
Set
$$
\th=e^{-bt/2}y(x,t).
$$
The \eq for $y$ is
$$
y_{tt}=\a^2y_{xx}+\frac {b^2}4 y
$$
with the same initial condition and the boundary condition
$$
y(0,t)=e^{bt/2}u(t).
$$

Apply the WKB method to find main singularities. The justification
can be found in \cite{BB}. Let
$$
u(t)=\dl(t).
$$
Write the solution as
$$
y(x,t)=\dl(t-x/a)+p(x)H(t-x/a) +q(x,t)(t-x/a)_+.
$$
Then the main terms of the LHS are
$$
\dl''+p\dl'+q\dl.
$$
And the main terms of the RHS are
$$
a^2\l[\frac1{a^2}\dl''+\frac1{a^2}p\dl'-2\frac1{a}p'\dl+\frac1{a^2}q\dl\r]
+\frac{b^2}4\dl.
$$
Matching the coefficients of the singularities we see that the
coefficients by $\dl''$ and $\dl'$ coincide for any $p$ and $q$.
Comparing the coefficients by $\dl$ we find
$$
2ap'=\frac{b^2}{4}.
$$
Taking into account the boundary condition we can write
$$
p(x)=\frac{b^2}{8}x.
$$
Now
$$
y(x,t)=\dl(t-x/a)+  \frac{b^2}{8a} x\,H(t-x/a) +q(x,t)(t-x/a)_+.
$$
And
$$
y(x,t)=e^{-bt/2}\dl(t-x/a)+  \frac{b^2}{8a}x e^{-bt/2}H(t-x/a) +q(x,t)e^{-bt/2}(t-x/a)_+.
$$
Because we are interested in the behavior near the front $x=at$,
this can be rewritten as
\EQn{
\th(x,t)=e^{-bx/2a}\dl(t-x/a)+  \frac{b^2}{8a} xe^{-bx/2a}H(t-x/a)+q_`(x,t)(t-x/a)_+.
}

In the case of \eqref K we can obtain simlar result with more complicated calculations.

\end{document}